%% file: main.tex
\documentclass[conference]{IEEEtran}

\usepackage{todonotes}
\usepackage{amsmath,amssymb,amsfonts}
\usepackage{algorithmic}
\usepackage[linesnumbered,ruled,vlined]{algorithm2e}
\usepackage{graphicx}
\usepackage{textcomp}
\usepackage{xcolor}
\usepackage{xspace}
\usepackage{hyperref}
\usepackage{microtype}
\usepackage{flushend}
\usepackage[T1]{fontenc}

\newcommand{\rnd}{\texttt{Random}\xspace}
\newcommand{\mosa}{\texttt{DynaMOSA}\xspace}
\newcommand{\nsga}{\texttt{NSGAII}\xspace}
\newcommand{\steady}{\texttt{Steady-State}\xspace}
\newcommand{\mupluslambda}{\texttt{Mu+Lambda}\xspace}

\newcommand{\evo}{\texttt{EvoSuite}\xspace}
\newcommand{\jjx}{\texttt{JoularJX}\xspace}

\newcommand{\rqa}{\textbf{RQ1}\xspace}
\newcommand{\rqab}{\textbf{RQ1.1}\xspace}
\newcommand{\rqb}{\textbf{RQ2}\xspace}

\begin{document}

\title{On the Energy Consumption of  Test Generation}

\author{

	\IEEEauthorblockN{Fitsum Kifetew}
	\IEEEauthorblockA{\textit{Software Engineering Unit} \\
		\textit{Fondazione Bruno Kessler}\\
		Trento, Italy \\
		kifetew@fbk.eu}
	\and
	\IEEEauthorblockN{Davide Prandi}
	\IEEEauthorblockA{\textit{Software Engineering Unit} \\
		\textit{Fondazione Bruno Kessler}\\
		Trento, Italy \\
		prandi@fbk.eu}
	\and
	\IEEEauthorblockN{Angelo Susi}
	\IEEEauthorblockA{\textit{Software Engineering Unit} \\
		\textit{Fondazione Bruno Kessler}\\
		Trento, Italy \\
		susi@fbk.eu}
	}

\maketitle

\begin{abstract}
Research in the area of automated test generation has seen remarkable progress in recent years, resulting in several approaches and tools for effective and efficient generation of test cases. In particular, the \evo tool has been at the forefront of this progress embodying various algorithms for automated test generation of Java programs. \evo has been used to generate test cases for a wide variety of programs as well. While there are a number of empirical studies that report results on the effectiveness, in terms of code coverage and other related metrics, of the various test generation strategies and algorithms implemented in \evo, there are no studies, to the best of our knowledge, on the energy consumption associated to the automated test generation. In this paper, we set out to investigate this aspect by measuring the energy consumed by \evo when generating tests. We also measure the energy consumed in the execution of the test cases generated, comparing them with those manually written by developers. The results show that the different test generation algorithms consumed different amounts of energy, in particular on classes with high cyclomatic complexity. Furthermore, we also observe that manual tests tend to consume more energy as compared to automatically generated tests, without necessarily achieving higher code coverage. Our results also give insight into the methods that consume significantly higher levels of energy, indicating potential points of improvement both for \evo as well as the different programs under test.   
\end{abstract}

\begin{IEEEkeywords}
	test generation, energy consumption
\end{IEEEkeywords}

\section{Introduction} \label{sec:introduction}
\input{sections/introduction}

\section{Background} \label{sec:background}
\input{sections/background}

\section{Experiment Design} \label{sec:experiment-design}
\input{sections/experimental_design}

\section{Results} \label{sec:results} 
\input{sections/results}

\section{Discussion} \label{sec:discussion}
\input{sections/discussion}

\section{Threats To Validity} \label{sec:validity}
\input{sections/validity}

\section{Conclusions and Future Works} \label{sec:conclusion}
\input{sections/conclusions}

\section*{Acknowledgment}
We acknowledge the support  of the PNRR project FAIR -  Future AI Research (PE00000013),  under the NRRP MUR program funded by the NextGenerationEU.

\bibliographystyle{IEEEtran}
\bibliography{biblio}

\end{document}

%% file: sections/introduction.tex
Automated test generation saves resources and helps reduce the cost of testing, eventually increasing the reliability of a piece of software. However, most automated test generation techniques adopt algorithms that are inherently based on search strategies, typically over an extremely large search space. This could make such techniques susceptible to potential inefficiency with respect to energy consumption. The energy inefficiency could be related to the test generation process itself or to the (repeated) execution of the generated test cases, say in the context of continuous integration. To shed light into this phenomenon, in this paper we analyze the energy consumption of automated test generation as well as those of the generated tests using the popular unit test generation tool \evo~\cite{DBLP:conf/sigsoft/FraserA11}. More specifically we aim to answer the following research questions: \\
\noindent
\textbf{\rqa (Test Generation Cost}): What is the energy consumption of the test generation process? Are there any differences among the various test generation algorithms? \\
\textbf{\rqab (Methods Profile}): How is energy consumption distributed at the method level? Which methods of which classes consume the most energy?
\\ \noindent
\textbf{\rqb (Test Execution Cost}): What is the energy consumption of executing test suites? 
    Is there any difference in energy consumption between automatically generated and  manually written test suites?

Earlier works (e.g., ~\cite{DBLP:journals/infsof/PanichellaKT18,DBLP:journals/infsof/CamposGAFEA18,DBLP:conf/ssbse/CamposGFEA17, DBLP:journals/tosem/FraserA14}) compared the various test generation algorithms in terms of their code coverage and fault finding ability (measured through mutation analysis) using the \evo tool. Works have also been carried out to assess the practical usefulness of automatically generated test cases through studies involving actual developers (e.g., ~\cite{DBLP:conf/issta/FraserSMAP13}) in which various issues of automatically generated tests have been reported. A recent work by Zaidman~\cite{zaidman2024inconvenient} studied the environmental impact of testing and continuous integration. However, to the best of our knowledge, there has not been any study that analyzed the energy consumption aspect of automated test generation and subsequent considerations.  

There are various methods for measuring the energy consumption of software, each offering different levels of detail and accuracy. Until recently, measuring energy consumption requires external metering hardware components, which implied a complex setup and did not provide fine-grained data~\cite{khan2018rapl}. 
An alternative is model based estimation, where a computational model maps performance reading (e.g., number of CPU cycles) into an estimate of the consumed energy. This approach does not require complicated hardware configurations, but its accuracy is quite limited~\cite{mccullough2011evaluating}. A third approach exploits energy measurement sensors installed directly on modern microprocessor architectures. For instance, Intel's Running Average Power Limit (RAPL)~\cite{intel2024intel} 
provides real-time energy usage of different components, such as the CPU cores or the integrated GPU. Several tools rely on RAPL to estimate energy consumption of software as, e.g., \cite{benoit_courty_2024_11171501,fieni2024powerapi,chattaraj2023rjoules, lella2024towards,chattaraj2023rjoules,DBLP:conf/intenv/Noureddine22}.
Among the other, we used the PowerJoular and \jjx tools, which are specifically designed for Java code~\cite{DBLP:conf/intenv/Noureddine22}. 
In particular, \jjx focuses on monitoring Java code at the method level providing a detailed report of the energy consumed by each method called by an application.

In order to address the research questions, we conducted an empirical study to measure the energy consumption of test case generation for five algorithms and more than 350 Java classes selected from a large set of projects~\cite{DBLP:conf/icse/GruberRPSM024} (\rqa and \rqab). For eight projects, we compared the energy consumption of generating and executing whole-project tests using the same algorithms (\rqb). Finally, for 162 projects, we executed tests manually written by developers as well as automatically generated (\rqb). 
The results of the study highlights marked differences in the energy consumption of the considered generation algorithms (\rqa).  
Moreover, the analysis of the methods (\rqab) reveals a significant variation in the energy contribution of each individual method.
The analysis of test execution indicates that the generation process requires significantly more energy than test execution (\rqb). Finally, comparing manually written and automatically generated tests shows that
complex manual test could be energy intensive without improving coverage compared to automatically generated tests \rqb).

The main contributions of this work are:
\begin{enumerate}
    \item Empirical evaluation of energy consumption: we conducted an extended empirical evaluation of the energy consumption of test generation algorithms implemented in \evo, focusing on Java programs. 
    We make available a replication package with experimental data as well as scripts for reproducing the results reported~\cite{replication_package}.
    \item Energy profile at the method level: we provided a complete energy characterization of the Java methods used during test generation, highlighting the methods that are more energy demanding.
    \item Test case execution cost: we measured the energy consumed by executing test cases. We evaluate the differences when considering test cases generated with different algorithms as well as the differences between manually written and generated test.
\end{enumerate}

The rest of the paper is organized as follows. Section~\ref{sec:background} introduces the basic concepts about the energy profile of software and automatic test generation. Section~\ref{sec:experiment-design} details the design of the experiments we performed. Section~\ref{sec:results} presents observed results and Section~\ref{sec:discussion} discusses their implications  Threats to validity are presented in Section~\ref{sec:validity} while Section~\ref{sec:conclusion} concludes the paper and outlines future work.

%% file: sections/background.tex
\subsection{Energy Profile of Sofware} \label{sec:EnergyProfileOfSoftware}
\input{sections/background-epf}

\subsection{Automatic Test Generation} \label{sec:AutomaticTestGeneration}
\input{sections/background-atg}

%% file: sections/background-epf.tex
Intel's Running Average Power Limit (RAPL)~\cite{intel2024intel}  is a power management and monitoring framework integrated into modern Intel processors. 
It provides real-time data on energy usage, allowing users to monitor power consumption for various hardware components.
RAPL uses  model specific registers (MSRs) updated approximately once every millisecond that report the energy consumed since the boot-up of the system. 
Depending on the specific hardware architecture, RAPL provides energy measurements for different power domain. A power domain represents a part of the hardware with its own power and energy characteristics. Modern CPU includes PSys domain that monitors energy consumption of the whole System on Chip (SoC), including all the cores, the integrated graphic, and the memory integrated on the chip.
The Linux Power Capping Framework\footnote{https://www.kernel.org/doc/html/latest/power/powercap/powercap.html} supplies a standard interface to read energy consumption data from RAPL registers, simplifying the development of energy measurement tools. 
The accuracy of the RAPL measurements has been evaluated in several studies~\cite{hackenberg2013power,huang2015measurement, desrochers2016validation, khan2018rapl} showing good concordance with energy measured at plug power.

Several tools based on Intel's RAPL have been developed to measure the energy usage of software. 
CodeCarbon~\cite{benoit_courty_2024_11171501} and pyJoules~\cite{fieni2024powerapi} are Python packages designed to estimate the power consumption of specific Python functions.  RJoules~\cite{chattaraj2023rjoules} focuses on the energy usage of R code snippets, while DBJoules~\cite{lella2024towards} allows for measuring the energy consumption of database queries.
Finally, the tools jRAPL~\cite{chattaraj2023rjoules} and \jjx~\cite{DBLP:conf/intenv/Noureddine22} are designed for use with Java code and are compatible with the \evo tool.
However, jRAPL requires the use of specific methods within the code to measure method-level energy consumption, making it less suitable for large-scale energy evaluation.
Instead, \jjx is designed as a JavaAgent that attaches to the Java Virtual Machine to monitor power consumption at runtime and can be run alongside \evo test case generation. 

\jjx is based on PowerJoular~\cite{DBLP:conf/intenv/Noureddine22}, a tool designed to monitor energy consumption of software components.
PowerJoular  tracks CPU and GPU energy consumption using the Intel’s RAPL interface.  \jjx builds on PowerJoular to monitor the source code
of Java applications. \jjx runs as a Java agent integrated within the Java Virtual Machine (JVM) allowing to  measure the power consumption of individual methods during execution. 
It uses PowerJoular to track overall CPU utilization and energy consumption, and statistically allocates this power usage to specific Java methods. 
Real-time tracking includes all Java methods, including those from the Java Development Kit (JDK) or other Java libraries, allowing a complete characterization of 
the energy footprint of a Java application. On Linux, \jjx uses the Linux Power Capping interface to read the available power domains and selects, when available, PSys domain.


%% file: sections/background-atg.tex
Automated Test Generation has come a long way over the years, in particular for Java applications, there are a number of mature tools and techniques proposed by the research community, as evidenced by the long running testing tool competitions~\cite{DBLP:journals/stvr/DevroeyGGJKPP23}. Notably \evo~\cite{DBLP:conf/sigsoft/FraserA11} is arguably the most popular test generation tool for Java, implementing a wide range of test generation criteria as well as assertion generation strategies. It includes a wide range of search algorithms for test generation~\cite{DBLP:journals/infsof/CamposGAFEA18,DBLP:journals/infsof/PanichellaKT18}. Given the nature of test generation algorithms, in particular those based on dynamically executing the program under test during the test generation process, the overall process is computationally intensive. While different aspects of the effectiveness and efficiency of the various tools and algorithms have been covered by different studies, aspects related to energy consumption of test generation have not so far been addressed. In this paper, we set out to investigate this issue with an empirical study on \evo and the various algorithms implemented in it. In this section, we briefly describe some details of the \evo tool which we believe are relevant for understanding the experiment we performed.

\evo has a multi-threaded client-server implementation where the \emph{client} process performs the actual test generation activity while the \emph{server} (master) process manages the overall process and tracks progress by continuously communicating with the client process. The master process also handles any communication with the tester by way of accepting command line arguments and finally reporting back the result of the test generation process. Hence, in our study related to the energy consumption of the automated test generation process, the roles of the client and master process are different. 
While the energy consumption due to the master process can easily be monitored by attaching the \jjx JavaAgent directly to the main Java Virtual Machine (JVM) when launching \evo, measuring the consumption due to the client process is not straightforward as it is not directly accessible. 
To be able to measure the energy consumption of the client process, we have modified \evo in such a way that it is possible to attach a generic JavaAgent to the client process (which otherwise is not reachable directly from outside). 

%% file: sections/experimental_design.tex
This section summarizes the empirical study we run to address the research questions we presented in Section~\ref{sec:experiment-design}. 

\subsection{Benchmark} \label{sec:Benchmark}

We performed our analysis starting from the large set of Java projects collected in a recent study~\cite{DBLP:conf/icse/GruberRPSM024}. 
These projects contain tests written by developers as well as generated tests. We utilized Lizard\footnote{https://github.com/terryyin/lizard} to extract the cyclomatic complexity (CCN)
for each method, and then calculated the CCN of a class as the average CCN  of its methods.
Overall, we collected code complexity data from 7389 classes in 350 projects. 

In order to estimate the energy consumption of the test generation process (\rqa), we selected two sets of Java classes: High CCN and Low CCN. High CCN consists of 73 classes from 42 projects and is defined as the set of classes in the top 1\% of the CCN distribution (mean CCN 16.4). Low CCN comprises 288 classes from 141 projects and were randomly selected (mean CCN 1.7). Therefore, the total number of considered Java classes is 361. 
The same sets are used to assess the profile of the energy consumption at the method level (\rqab).

To  evaluate the energy costs of test execution (\rqb), we would have ideally used the tests generated for answering \rqa and \rqab. 
However, our analysis of \rqa and \rqab focused on classes that resulted in small tests, and the energy consumed by their execution turned out to be too low for \jjx to measure accurately.  Therefore, to address \rqb, we instead considered eight randomly selected projects, which included a total of 350 classes. The number of classes in the eight projects is comparable with the number of classes used for \rqa and \rqab. However, note that the two sets of classes are different.

Finally, to compare the energy consumption of automatically generated and manually written test suites (\rqb), we executed both manually written and generated tests included in the original Java projects. We were able to compile and execute manual and generated tests for 162 projects. The main reasons for failure are the lack of dependencies and the requirement of a legacy Java version (e.g., JDK version 6). Note that tests are generated with the \evo tool. 

\subsection{Empirical Procedure}

For comparing the energy consumption of different test generation algorithms (\rqa), we considered, among the various implemented into the \evo tool, five of them representative of four broader categories:
\begin{itemize}
    \item Random search: \rnd
    \item Genetic algorithm: \steady
    \item Evolutionary strategies: \mupluslambda
    \item Multi/many objectives:  \nsga and \mosa
\end{itemize}
Furthermore, because \evo has a client-server architecture (see Section~\ref{sec:AutomaticTestGeneration}), we only measured client energy to reduce \evo-specific energy consumption. Additionally, we have disabled \evo post processing for the same purpose. We then ran \evo test case generation along with \jjx JavaAgent  30 times for each algorithm and for each class in  High and Low CCN sets, resulting in a total of 10,530 runs. All test case generation algorithms were executed using the same time budget of 180 seconds, while the other \evo parameters were configured with default values.

The same methodology assessed test execution costs (\rqb). For the eight selected Java projects, we ran \evo tool along with \jjx JavaAgent 10 times\footnote{To lower the computational burden of generating whole project tests, we used 10 replicas instead of 30. The entire process required more than 12 days.} for each algorithm, resulting in 166,407 tests. In this experiment, we enabled \evo post processing to store on the disk the generated tests. We subsequently executed these tests 30 times using JUnit~\cite{tudose2020junit} and \jjx JavaAgent to measure energy consumption during test execution. This experiment allowed us to compare the energy footprint of test generation and execution.

The last experiment we conducted aimed to understand the differences, if any, between manually and automatically generated tests (\rqb). For each project, we executed  manual and automatically generated tests 30 times using JUnit with the \jjx JavaAgent. 

All the experiments are performed on a server equipped with Intel Xeon Gold 6226R with 196GB of RAM. This CPU has the Cascade Lake architecture that supports PSys energy domain. The operating system used was Ubuntu 20.04.1 LTS, and \jjx reads the Psys domain through the Linux Power Capping Framework (Section~\ref{sec:EnergyProfileOfSoftware}). 

\subsection{Performance metrics}

Both the \evo test generation process and \jjx energy measurements are stochastic processes, meaning their outputs vary from run to run. The empirical procedure outlined in the previous section involves performing multiple repetitions to address statistical fluctuations. When we report the consumption data of an algorithm for a Java class or project, we are referring to the average energy consumption across these repetitions. Furthermore, when comparing two energy distributions, we use the Wilcoxon rank sum test followed by Benjamini-Hochberg procedure~\cite{benjamini1995controlling} with a significance level of 5\% to control the false discovery rate (FDR). When applicable, we compute effect size using the Vargha-Delaney statistic (\^{A}).

%% file: sections/results.tex
In this section, we present the results of our experiment and provide answers to the research questions outlined earlier. First, we look at the energy consumption of the test generation process (\rqa), followed by an examination of the energy consumption due to the execution of test suites, of those automatically generated as well as those manually written by developers (\rqb).

\subsection{\rqa: Generation Cost}
\begin{figure}[tb]
\includegraphics[width=\linewidth]{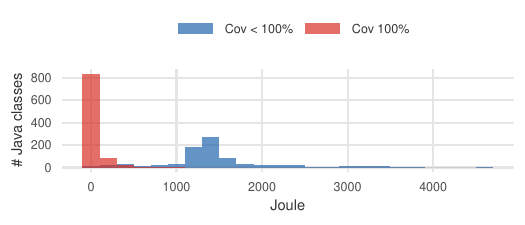}
\caption{Energy consumption stratified by full coverage  achievement. The y axis represents the number (\#) of classes. Note that data is not stratified per generation algorithm, and therefore classes are counted five times, one for
each algorithm considered.}
\label{fig:energy_hist}
\end{figure}

With \rqa, we aim to investigate how much energy is consumed when generating test cases for Java programs, and whether or not the energy consumed varies with the test generation algorithm used. Furthermore, we also explore whether or not the energy consumed is dependent on the complexity of the system under test (CCN) and whether some methods/packages lead to higher energy consumption. Hence, we first take a look at the energy consumed during test generation by categorizing the systems under test into those where full coverage was achieved and those where full coverage was not achieved. Figure~\ref{fig:energy_hist} shows a density plot (histogram) of the total energy stratified by full coverage. 

As can be seen from Figure~\ref{fig:energy_hist}, the energy consumed during test generation tends to be much lower when full coverage is achieved, while higher consumption is observed when test generation fails to achieve full coverage, i.e.,  the algorithm runs for the full duration of the search budget. 
This later could potentially be because the system under test exhibits high complexity, making test generation difficult. 

\begin{figure*}[tb]
\includegraphics[width=\textwidth]{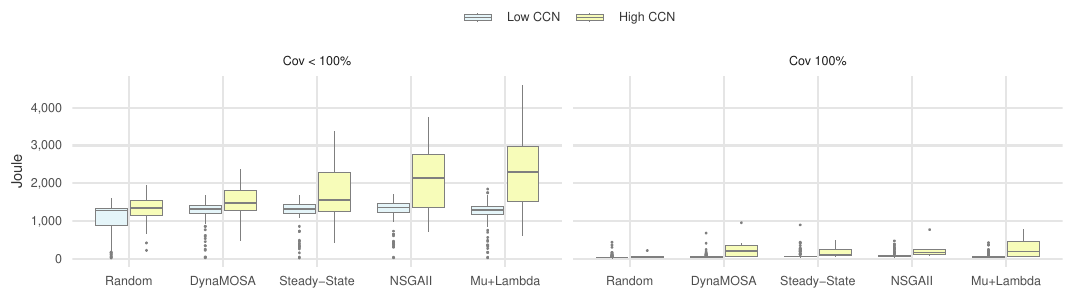}
\caption{Boxplot of energy consumption per algorithm stratified by CCN and full coverage achievement. } 
\label{fig:energy_boxplot}
\end{figure*}

To get further insight into the factors that play into the overall energy consumption of test generation, we consider the CCN of the systems under test by categorizing them into two groups: \emph{Low CCN} and \emph{High CCN} (Section~\ref{sec:Benchmark}). Figure~\ref{fig:energy_boxplot} reports the boxplots of the energy consumption of each search algorithm on subjects with low and high CCN, when full coverage is not achieved (figure on the left) and when it is (figure on the right). 

As we observed in Figure~\ref{fig:energy_hist} as well, we also see in Figure~\ref{fig:energy_boxplot} that for subjects where full coverage is achieved, the energy consumed is lower, regardless of the algorithm.
We also observe that the energy consumed tends to be higher for subjects with high CCN, independently of the algorithm and the fact that full coverage was achieved or not. Looking closer at Figure~\ref{fig:energy_boxplot} we notice that when full coverage is not achieved, \mupluslambda tends to consume higher, followed respectively by \nsga and \steady. However, when full coverage is achieved (figure to the right), \mosa tends to consume higher as well, though the difference is not statistically significant. We also observe interestingly that \rnd tends to consume less energy in both cases with high CCN.

To support Figure~\ref{fig:energy_boxplot} with statistical tests of significance, we present detailed comparisons in Table~\ref{tab:energy_boxplot} where we can see the result of applying the Wilcoxon test, adjusted using the Benjamini-Hochberg procedure to control FDR. Each line of the table compares the energy consumption of two sets of Java classes. For instance, line 1 compares the energy consumption of test suite generation between Low CCN (Group 1) and High CCN (Group 2) classes when using \rnd (Aggregate) and not achieving full coverage (Coverage). The energy consumption for Low CCN classes (Energy G1) is less than that of High CCN classes (Energy G2), with a small effect size (Effect size (\^{A})) and a significant adjusted p-value (p\_adjusted).
Moreover, to facilitate easy reading, we have numbered and separated the rows in the table into three sections by double lines, and within each section we further make divisions using single lines to mark the different subcategories. For instance, we see from Table~\ref{tab:energy_boxplot} (rows 1-10) that the difference between \emph{Low CCN} and \emph{High CCN} is statistically significant, independently of the search algorithm used and whether or not full coverage was achieved. 

The second and third sections of Table~\ref{tab:energy_boxplot} present pairwise comparisons among the algorithms when full coverage is not achieved (rows 11-30) and when full coverage is achieved (rows 31-50). For instance, when full coverage is not achieved for classes with low CCN (rows 11-20), differences among the algorithms are not statistically significant, except for \rnd vs \nsga. On the other hand, looking at classes with high CCN where coverage is less than 100\% (rows 21-30), we notice that all differences among the algorithms are significant, except for those between \mosa and \steady, \mupluslambda and \nsga, and \nsga and \steady. Similarly for classes with low CCN where full coverage is achieved (rows 31-40) we see that all differences among the algorithms are statistically significant, while for classes with high CCN (rows 41-50), none of the differences are statistically significant.

\begin{table*}[htb]
\centering
\begingroup\fontsize{7pt}{7pt}\selectfont
\begin{tabular}{lllllrrlrr}
  \hline
No. & Group 1 & Group 2 & Aggregate & Coverage & Energy G1. (J) & Energy G2. (J) & Effect size (\^{A}) & pval & p\_adjusted \\ 
  \hline
1 & Low CCN & High CCN & Random & $<$100\% & 1,283.890 & 1,355.880 & 0.364  ( small ) & 0.003 & \textbf{0.006} \\ 
  2 & Low CCN & High CCN & DynaMOSA & $<$100\% & 1,328.570 & 1,473.130 & 0.334  ( medium ) & 0.000 & \textbf{0.001} \\ 
  3 & Low CCN & High CCN & Mu+Lambda & $<$100\% & 1,293.650 & 2,302.620 & 0.078  ( large ) & 0.000 & \textbf{0.000} \\ 
  4 & Low CCN & High CCN & NSGAII & $<$100\% & 1,356.580 & 2,176.080 & 0.217  ( large ) & 0.000 & \textbf{0.000} \\ 
  5 & Low CCN & High CCN & Steady-State & $<$100\% & 1,332.780 & 1,749.530 & 0.288  ( medium ) & 0.000 & \textbf{0.000} \\ \hline
  6 & Low CCN & High CCN & DynaMOSA & 100\% & 45.790 & 205.700 & 0.119  ( large ) & 0.000 & \textbf{0.000} \\ 
  7 & Low CCN & High CCN & Mu+Lambda & 100\% & 39.870 & 195.370 & 0.099  ( large ) & 0.000 & \textbf{0.000} \\ 
  8 & Low CCN & High CCN & NSGAII & 100\% & 64.040 & 165.820 & 0.174  ( large ) & 0.007 & \textbf{0.013} \\ 
  9 & Low CCN & High CCN & Random & 100\% & 36.350 & 54.090 & 0.193  ( large ) & 0.020 & \textbf{0.036} \\ 
  10 & Low CCN & High CCN & Steady-State & 100\% & 59.340 & 97.430 & 0.178  ( large ) & 0.007 & \textbf{0.014} \\ \hline \hline
  11 & Random & DynaMOSA & Low CCN & $<$100\% & 1,283.890 & 1,328.570 & 0.423  ( small ) & 0.066 & 0.100 \\ 
  12 & Random & Mu+Lambda & Low CCN & $<$100\% & 1,283.890 & 1,293.650 & 0.483  ( negligible ) & 0.680 & 0.765 \\ 
  13 & Random & NSGAII & Low CCN & $<$100\% & 1,283.890 & 1,356.580 & 0.375  ( small ) & 0.002 & \textbf{0.005} \\ 
  14 & Random & Steady-State & Low CCN & $<$100\% & 1,283.890 & 1,332.780 & 0.426  ( small ) & 0.076 & 0.112 \\ 
  15 & Mu+Lambda & DynaMOSA & Low CCN & $<$100\% & 1,293.650 & 1,328.570 & 0.55  ( negligible ) & 0.234 & 0.300 \\ 
  16 & DynaMOSA & NSGAII & Low CCN & $<$100\% & 1,328.570 & 1,356.580 & 0.455  ( negligible ) & 0.288 & 0.343 \\ 
  17 & DynaMOSA & Steady-State & Low CCN & $<$100\% & 1,328.570 & 1,332.780 & 0.502  ( negligible ) & 0.958 & 0.998 \\ 
  18 & Mu+Lambda & NSGAII & Low CCN & $<$100\% & 1,293.650 & 1,356.580 & 0.41  ( small ) & 0.032 & 0.055 \\ 
  19 & Mu+Lambda & Steady-State & Low CCN & $<$100\% & 1,293.650 & 1,332.780 & 0.452  ( negligible ) & 0.254 & 0.318 \\ 
  20 & Steady-State & NSGAII & Low CCN & $<$100\% & 1,332.780 & 1,356.580 & 0.547  ( negligible ) & 0.264 & 0.322 \\ \hline
  21 & DynaMOSA & Mu+Lambda & High CCN & $<$100\% & 1,473.130 & 2,302.620 & 0.206  ( large ) & 0.000 & \textbf{0.000} \\ 
  22 & DynaMOSA & NSGAII & High CCN & $<$100\% & 1,473.130 & 2,176.080 & 0.303  ( medium ) & 0.000 & \textbf{0.000} \\ 
  23 & Random & DynaMOSA & High CCN & $<$100\% & 1,355.880 & 1,473.130 & 0.614  ( small ) & 0.024 & \textbf{0.043} \\ 
  24 & DynaMOSA & Steady-State & High CCN & $<$100\% & 1,473.130 & 1,749.530 & 0.398  ( small ) & 0.045 & 0.072 \\ 
  25 & NSGAII & Mu+Lambda & High CCN & $<$100\% & 2,176.080 & 2,302.620 & 0.575  ( small ) & 0.141 & 0.186 \\ 
  26 & Random & Mu+Lambda & High CCN & $<$100\% & 1,355.880 & 2,302.620 & 0.858  ( large ) & 0.000 & \textbf{0.000} \\ 
  27 & Steady-State & Mu+Lambda & High CCN & $<$100\% & 1,749.530 & 2,302.620 & 0.683  ( medium ) & 0.000 & \textbf{0.001} \\ 
  28 & Random & NSGAII & High CCN & $<$100\% & 1,355.880 & 2,176.080 & 0.771  ( large ) & 0.000 & \textbf{0.000} \\ 
  29 & Steady-State & NSGAII & High CCN & $<$100\% & 1,749.530 & 2,176.080 & 0.603  ( small ) & 0.039 & 0.066 \\ 
  30 & Random & Steady-State & High CCN & $<$100\% & 1,355.880 & 1,749.530 & 0.313  ( medium ) & 0.000 & \textbf{0.001} \\ \hline \hline
  31 & Mu+Lambda & DynaMOSA & Low CCN & 100\% & 39.870 & 45.790 & 0.612  ( small ) & 0.000 & \textbf{0.000} \\ 
  32 & DynaMOSA & NSGAII & Low CCN & 100\% & 45.790 & 64.040 & 0.188  ( large ) & 0.000 & \textbf{0.000} \\ 
  33 & Random & DynaMOSA & Low CCN & 100\% & 36.350 & 45.790 & 0.711  ( medium ) & 0.000 & \textbf{0.000} \\ 
  34 & DynaMOSA & Steady-State & Low CCN & 100\% & 45.790 & 59.340 & 0.241  ( large ) & 0.000 & \textbf{0.000} \\ 
  35 & Mu+Lambda & NSGAII & Low CCN & 100\% & 39.870 & 64.040 & 0.161  ( large ) & 0.000 & \textbf{0.000} \\ 
  36 & Random & Mu+Lambda & Low CCN & 100\% & 36.350 & 39.870 & 0.601  ( small ) & 0.001 & \textbf{0.002} \\ 
  37 & Mu+Lambda & Steady-State & Low CCN & 100\% & 39.870 & 59.340 & 0.194  ( large ) & 0.000 & \textbf{0.000} \\ 
  38 & Random & NSGAII & Low CCN & 100\% & 36.350 & 64.040 & 0.878  ( large ) & 0.000 & \textbf{0.000} \\ 
  39 & Steady-State & NSGAII & Low CCN & 100\% & 59.340 & 64.040 & 0.6  ( small ) & 0.001 & \textbf{0.002} \\ 
  40 & Random & Steady-State & Low CCN & 100\% & 36.350 & 59.340 & 0.145  ( large ) & 0.000 & \textbf{0.000} \\ \hline
  41 & Mu+Lambda & DynaMOSA & High CCN & 100\% & 195.370 & 205.700 & 0.478  ( negligible ) & 0.903 & 0.960 \\ 
  42 & NSGAII & DynaMOSA & High CCN & 100\% & 165.820 & 205.700 & 0.5  ( negligible ) & 1.000 & 1.000 \\ 
  43 & Random & DynaMOSA & High CCN & 100\% & 54.090 & 205.700 & 0.778  ( large ) & 0.110 & 0.148 \\ 
  44 & Steady-State & DynaMOSA & High CCN & 100\% & 97.430 & 205.700 & 0.537  ( negligible ) & 0.860 & 0.934 \\ 
  45 & NSGAII & Mu+Lambda & High CCN & 100\% & 165.820 & 195.370 & 0.5  ( negligible ) & 1.000 & 1.000 \\ 
  46 & Random & Mu+Lambda & High CCN & 100\% & 54.090 & 195.370 & 0.82  ( large ) & 0.058 & 0.090 \\ 
  47 & Steady-State & Mu+Lambda & High CCN & 100\% & 97.430 & 195.370 & 0.583  ( small ) & 0.625 & 0.727 \\ 
  48 & Random & NSGAII & High CCN & 100\% & 54.090 & 165.820 & 0.833  ( large ) & 0.083 & 0.115 \\ 
  49 & Steady-State & NSGAII & High CCN & 100\% & 97.430 & 165.820 & 0.583  ( small ) & 0.689 & 0.765 \\ 
  50 & Random & Steady-State & High CCN & 100\% & 54.090 & 97.430 & 0.167  ( large ) & 0.083 & 0.115 \\ 
   \hline
\end{tabular}
\endgroup
\caption{Statistical test of significance for the results in Fig.~\ref{fig:energy_boxplot} using Wilcoxon, with p-values adjusted for FDR (column \emph{p\_adjusted} shows the adjusted p values). Effect size computed using the Vargha-Delaney statistic (\^{A}). Significant p-values are in boldface.}
\label{tab:energy_boxplot}
\end{table*}

\subsection{\rqab: Methods Profile} \label{sec:methods_profile}
Taking a closer look, considering the energy consumption data for each test generation run, we inspect the methods in the top 5\%. Table~\ref{tab:top_consuming_methods_freq} presents the methods at the top 5\% of at least 20 classes in the systems under test. Table~\ref{tab:top_consuming_methods_freq} reports, for each method, in how many classes (column \emph{No. classes}) the method appears in the top 5\% of the most consuming methods, regardless of the search algorithm used. The table also reports for each algorithm in how many classes the given method was in the top 5\%. For instance, the method \texttt{java.lang.Thread.dumpThreads} was in the top 5\% for 355 classes overall, and for 341 classes where the algorithm was \nsga, 343 classes where the algorithm was \mosa, etc. The results show that most of the top consuming methods are those with low level functionality that are likely to be called repeatedly from different high level methods. We also notice that most of them are from libraries used by \evo, either provided by Java or third parties. Only three methods in the list are from \evo (names shown in boldface in Table~\ref{tab:top_consuming_methods_freq}).

\begin{table*}[tb]
\centering
\begingroup\fontsize{6pt}{7pt}\selectfont
\begin{tabular}{lrrrrrr} 
\hline
 Method name & No. classes & \nsga & \mosa & \mupluslambda & \rnd & \steady \\ 
  \hline
java.lang.Thread.dumpThreads & 355 & 341 & 343 & 347 & 334 & 346 \\ 
  jdk.internal.misc.Unsafe.unpark & 251 & 201 & 144 & 90 & 32 & 189 \\ 
  java.net.SocketInputStream.socketRead0 & 182 & 165 & 159 & 174 & 178 & 165 \\ 
  sun.nio.cs.UTF\_8\$Decoder.decodeArrayLoop & 176 & 158 & 128 & 124 & 113 & 141 \\ 
  java.io.RandomAccessFile.readBytes & 174 & 121 & 105 & 96 & 103 & 121 \\ 
  java.lang.ClassLoader.defineClass1 & 164 & 161 & 161 & 158 & 151 & 159 \\ 
  java.util.HashMap.hash & 157 & 136 & 111 & 104 & 108 & 124 \\ 
  java.util.zip.Inflater.inflateBytesBytes & 149 & 143 & 132 & 119 & 101 & 140 \\ 
  io.github.xstream.mxparser.MXParser.nextImpl & 148 & 140 & 139 & 142 & 139 & 141 \\ 
  jdk.internal.misc.Unsafe.park & 126 & 67 & 97 & 85 & 33 & 15 \\ 
  java.util.HashMap.putVal & 116 & 101 & 15 & 1 & 6 & 84 \\ 
  \textbf{org.evosuite.testcase.variable.VariableReferenceImpl.getStPosition} & 104 & 86 & 1 & 0 & 47 & 14 \\ 
  java.lang.Class.getEnclosingMethod0 & 51 & 36 & 1 & 3 & 21 & 3 \\ 
  io.github.xstream.mxparser.MXParser.parseStartTag & 49 & 30 & 24 & 25 & 24 & 32 \\ 
  java.lang.StackTraceElement.initStackTraceElements & 42 & 25 & 11 & 14 & 6 & 25 \\ 
  \textbf{org.evosuite.utils.ListenableList\$ObservableListIterator.hasNext} & 38 & 29 & 0 & 0 & 24 & 0 \\ 
  java.lang.Object.clone & 35 & 33 & 0 & 0 & 1 & 25 \\ 
  \textbf{org.evosuite.testcase.TestChromosome.size} & 32 & 0 & 0 & 0 & 0 & 32 \\ 
  java.net.SocketOutputStream.socketWrite0 & 30 & 1 & 0 & 23 & 0 & 13 \\ 
  java.lang.StringBuilder.append & 22 & 17 & 2 & 1 & 2 & 18 \\ 
  java.lang.Class.getDeclaringClass0 & 22 & 15 & 1 & 1 & 10 & 1 \\  \hline
  \end{tabular}
\endgroup
\caption{The methods with high energy consumption (top 5\%) in each class. Methods belonging to \evo shown in boldface.}
\label{tab:top_consuming_methods_freq}
\end{table*}

\begin{figure*}[t]
\includegraphics[width=\textwidth]{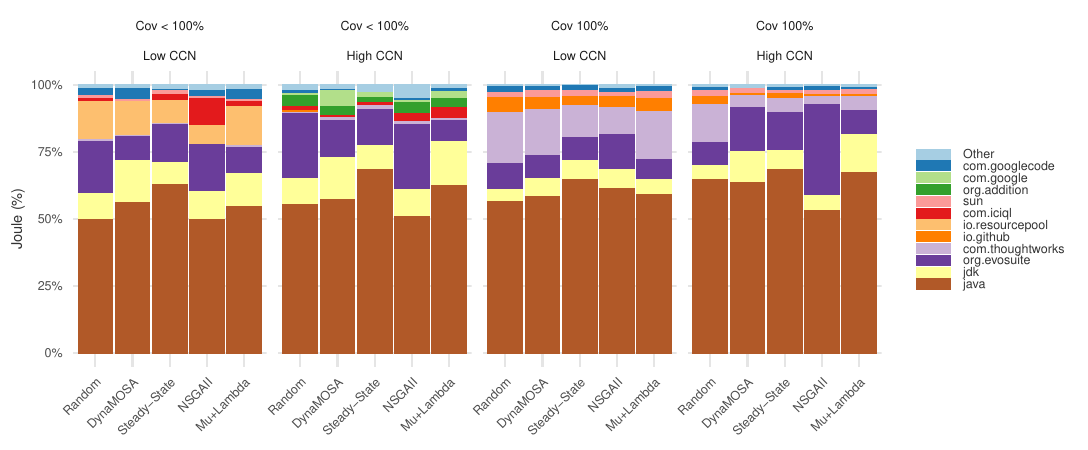}
\caption{Barplot of methods energy consumption per algorithm stratified by CCN and full coverage achievement.}
\label{fig:energy_method_barplot}
\end{figure*}

Zooming out from the specific methods and looking at the high level packages to which these methods belong, we obtain the barplot in Figure~\ref{fig:energy_method_barplot} where we see, for each algorithm, the packages of the top consuming methods. The data is stratified in a similar manner as previous plots considering CCN and full coverage status.

\subsection{\rqb: Tests Execution Cost} \label{sec:test_execution_cost}

In \rqb we investigate the energy consumption of executing test suites. Typically tests are generated/written once, and executed several times (e.g., in a continuous integration setting). Hence, the energy consumption of the execution of the test suites could be a relevant aspect to take into consideration. To get insight into this aspect, we look at the energy consumption of executing the test suites generated by the various algorithms for eight projects (Section~\ref{sec:Benchmark}). We also compare the energy consumption of executing manually written test suites with that of automatically generated ones.

\begin{figure*}[tb]
\includegraphics[width=\textwidth]{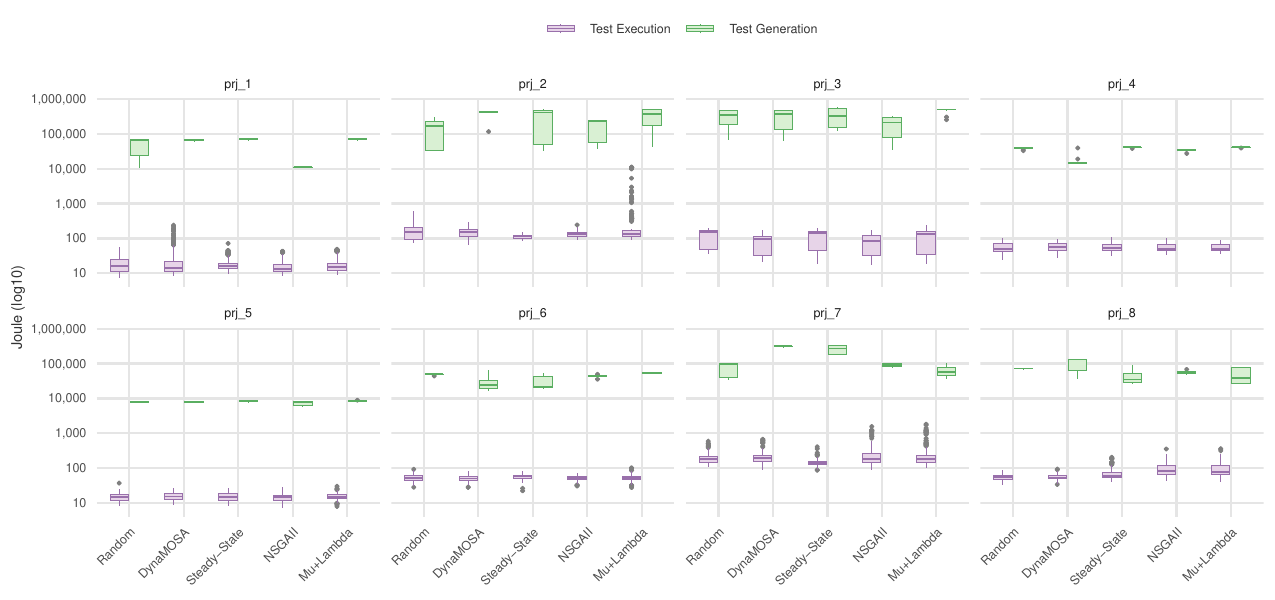}
\caption{Boxplot of the energy consumption of  whole project test case generation (above) and execution (below). The y axis reports the $log_{10}$ of the Joule consumed.  
Projects names: prj\_1 = 15058126273-mysql-table-helper, prj\_2 = alipay-sofa-common-tools, prj\_3 = amazonwebservices-aws-elasticache-cluster-client-memcached-for-java, 
                      prj\_4 = carlosthe19916-sunat-web-services, prj\_5 = CatDou-common-export, prj\_6 = chrlembeck-chrlembeck-util, prj\_7 = codelibs-jhighlight, prj\_8 = easyinnova-Tiff-Library-4J}
\label{fig:generate_tests_per_project_summary_boxplot}
\end{figure*}

Figure~\ref{fig:generate_tests_per_project_summary_boxplot} shows the energy consumption during the execution of test suites generated automatically by the different algorithms\footnote{here we present data for eight randomly selected projects to limit the number of executions we had to perform}. To give perspective, the figure also shows the energy consumption of generating tests. As we can see from the figure, there does not appear to be a huge difference in the energy consumption of executing the test suites generated by the different algorithms. Furthermore, aside from some outliers (in particular for \mupluslambda), the energy consumed is generally low. On the other hand, looking at the energy consumption of the test generation process for the same set of systems under test using the different algorithms, we notice that there are indeed differences among the generation algorithms. We also notice that test generation consumes much higher levels of energy compared to test execution.

\begin{figure*}[tb]
\includegraphics[width=\textwidth]{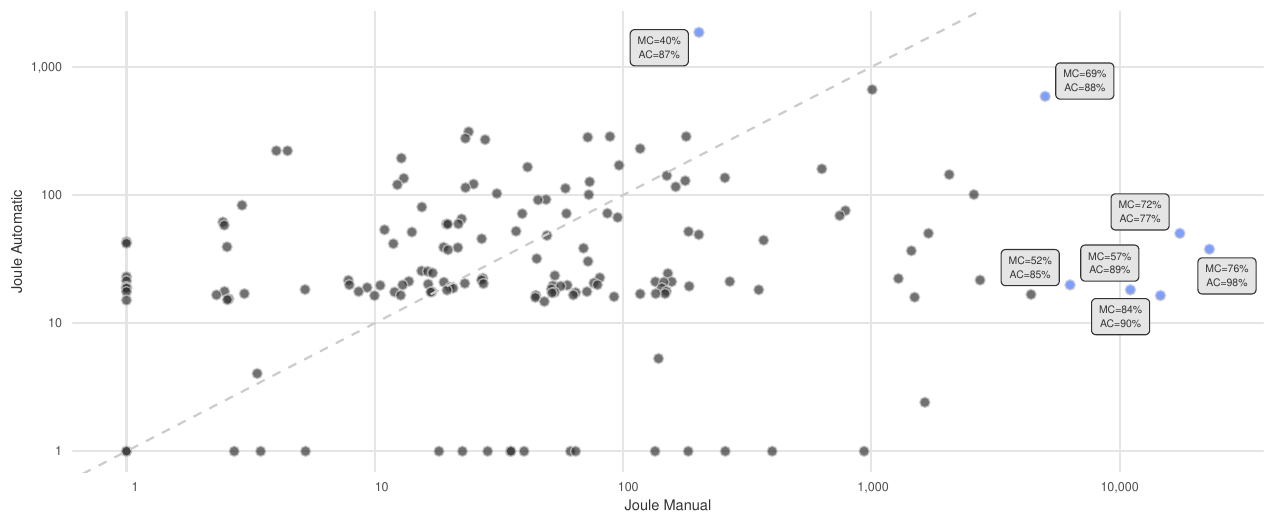}
\caption{Scatter plot comparing the energy consumption of executing manually written (x-axis) and automatically generated (y-axis) tests for 162 projects. Each dot represents a project and its median energy consumption measured with \jjx over 30 replicas. Selected outliers (blue points) are annotated with the branch coverage reached by manual (MC) and automatic (AC) tests. Note that the x and y axis are given in the $\log_{10}$ scale.}
\label{fig:manual_vs_auto_dotplot}
\end{figure*}

Finally, we compare the energy consumption of executing test suites manually written by developers with those automatically generated. In Figure~\ref{fig:manual_vs_auto_dotplot} we plot, for all the 162 projects in the dataset~\cite{DBLP:conf/icse/GruberRPSM024} (Section~\ref{sec:Benchmark}), the energy consumption of manually written (x-axis) against that of automatically generated test suites (y-axis). For this analysis, we executed all the test suites (both manually written by developers of the respective projects as well as those generated by \evo) present in the original dataset used for our experiments. That is, we did not generate the test suites as they were already available in the dataset, and for this particular analysis, we deemed it sufficient to use the test suites generated by the default \evo configurations available in the dataset. As can be seen from Figure~\ref{fig:manual_vs_auto_dotplot}, most of the test suites referred to each one of the projects (both manually written and generated automatically) have low energy consumption (mostly below a few hundred Joules) with a few exceptions where the consumption is much higher. For these outlier cases (shown in Figure~\ref{fig:manual_vs_auto_dotplot} in light blue color) we computed the branch coverage\footnote{using JaCoCo \url{https://www.jacoco.org/jacoco/}} achieved by the test suites as well as the complexity of the test code (as described in Section~\ref{sec:Benchmark}) to get further insight. It turns out that the manually written test suites exhibit a high level of code complexity (1.36-4.76 on average, maximum 135) compared to those automatically generated (1.16-1.45 on average, maximum 8). Looking at the branch coverage achieved (shown in gray boxes next to the blue circles in Figure~\ref{fig:manual_vs_auto_dotplot}), we notice that the execution of manually written test suites tends to consume high levels of energy while achieving lower levels of coverage as compared to automatically generated test suites.
 
%% file: sections/discussion.tex
The study presented in this paper is exploratory in nature with the main objective of getting insight into the energy consumption aspect of test generation, as such we did not set out to prove or disprove hypotheses but rather to quantify the energy consumed by test generation and eventually the execution of the test suites. While in principle any test generation tool could be used for such a study, we opted for \evo as it is readily available, it implements different test generation algorithms, and it is the most widely used tool in the automated testing literature. \evo however has specific design choices that should be taken into consideration with respect to energy consumption, the most important of which is that \evo separates the test generation process into \emph{master} and \emph{client} processes that run in parallel, with continuous communication for exchanging data. The actual test generation algorithm runs in the \emph{client} process. In our experiments we measure the energy consumption of both processes, however, we reported the results (Section~\ref{sec:results}) for the \emph{client} process only. This is because we wanted to focus on the actual test generation process, which allows us to compare the different algorithms. The \emph{master} process does mainly housekeeping activities such as keeping track of statistics, writing out tests and reports, etc. In our experiments, the \emph{master} process accounts on average for 22\% of the total energy consumed during the test generation process. Furthermore, the relative consumption of the \emph{master} process is higher when full coverage is achieved. Specifically, the \emph{master} process accounts for 41\% of the total consumption when full coverage is achieved, while for cases where full coverage was not achieved the proportion drops to 17\%. In other words, the longer the search algorithm runs, the less the relative consumption by the \emph{master} process.

Another aspect to take into account regarding the top consuming methods (Section~\ref{sec:methods_profile}) is that handling of threads, eventually resulting in dumping of threads, which leads to high energy consumption, as can be seen from Table~\ref{tab:top_consuming_methods_freq}. In the case of \evo, test execution heavily relies on threads, eventually resulting in frequent calls to the method \texttt{java.lang.Thread.dumpThreads}. An example of such call sequences is the following: \\

\noindent\texttt{TestCaseExecutor.execute} $\rightarrow$ \\ 
\texttt{org.evosuite.testcase.execution.}  \\
\hspace*{2mm} \texttt{.TestRunnable.storeCurrentThreads} $\rightarrow$ \\
\texttt{org.evosuite.runtime.thread.} \\ 
\hspace*{2mm} \texttt{.ThreadStopper.storeCurrentThreads} $\rightarrow$ \\
\texttt{java.lang.Thread.getAllStackTraces} $\rightarrow$ \\
\texttt{java.lang.Thread.dumpThreads} \\

Full details of the energy consumption information with the specific call sequences together with all the other experimental data are available in our replication package~\cite{replication_package} for further inspection.

Coming back to the different algorithms and their energy consumption, the results show that the principal factor impacting energy consumption is the code complexity of the systems under test, independent of whether or not full coverage was achieved. This is true for all the algorithms in our experiments. It is also true that when full coverage is achieved, the differences in the energy consumption of the different algorithms are minimal. While, when full coverage is not achieved, there are observable differences among the algorithms as shown in Figure~\ref{fig:energy_boxplot}. However, Figure~\ref{fig:energy_boxplot} does not show the full picture. In particular, it does not show the number of times an algorithm performs better than another in terms of coverage. To this end, we have analyzed the total dominance relation among the algorithms, i.e., the proportion of classes for which one algorithm dominates another (performs statistically significantly better both in terms of code coverage and energy consumption). As an example, in Tables~\ref{tab:pairwise_dynamosa} and~\ref{tab:pairwise_mulambda} we show the dominance tables for \mosa and \mupluslambda respectively. Due to lack of space, we do not report the tables for the other algorithms, however, the full data is available in the replication package~\cite{replication_package}. The dominance table for an algorithm reports the percentage of classes where the algorithm totally dominates the other algorithms (\emph{Total Dominance} column), and dominates with respect to energy consumption (\emph{Energy Dominance} column) when there is no statistically significant difference in coverage. For example, \mosa totally dominates (i.e., achieves better coverage and consumes less energy) \mupluslambda in 2\% of the classes, \nsga in 9\%, \rnd in 4\% and \steady in 7\%. It is worth noting that \mupluslambda has a higher energy dominance over \mosa, even though \mosa has (slightly) higher total dominance. 

\begin{table}[tb]
\centering
\begingroup\fontsize{6pt}{7pt}\selectfont
\begin{tabular}{lrr}
  \hline
Algorithm & Total Dominance (\%) & Energy Dominance (\%)  \\ 
  \hline
Mu+Lambda & 2 & 21  \\ 
  NSGAII & 9 & 75  \\ 
  Random & 4 & 9 \\ 
  Steady-State & 7 & 68  \\ 
   \hline
\end{tabular}
\endgroup
\caption{Dominance table for \mosa}
\label{tab:pairwise_dynamosa}
\end{table}

\begin{table}[tb]
\centering
\begingroup\fontsize{6pt}{7pt}\selectfont
\begin{tabular}{lrr}
  \hline
Algorithm & Total Dominance (\%) & Energy Dominance (\%) \\ 
  \hline
Dynamosa & 1 & 57 \\ 
  NSGAII & 5 & 68  \\ 
  Random & 4 & 11  \\ 
  Steady-State & 5 & 64  \\ 
   \hline
\end{tabular}
\endgroup
\caption{Dominance table for \mupluslambda}
\label{tab:pairwise_mulambda}
\end{table}

Looking at the energy consumption of test suite execution (Section~\ref{sec:test_execution_cost}), one aspect worth mentioning is that in some cases we have noticed manually written test cases that are thousands of lines long and have quite high cyclomatic complexity. As an example, in the project \texttt{albertoirurueta-irurueta-navigation-indoor} (one of the high consuming projects in Figure~\ref{fig:manual_vs_auto_dotplot}), the test \texttt{BeaconWithPowerAndLocated3DTest. .testConstructor}\footnote{\url{https://github.com/albertoirurueta/irurueta-navigation-indoor/blob/master/src/test/java/com/irurueta/navigation/indoor/BeaconWithPowerAndLocated3DTest.java}} is more than 2600 lines long with hundreds of assertions and exception handling constructs, leading to a high CCN of 135. The high CCN is due to a high number of exception testing constructs (try-catch-fail) in the same test, which increases the number of independent paths in the test. This mechanism is also used by \evo for exception testing, however, the number of such constructs in \evo generated tests is quite low, hence we do not observe high CCN values for tests generated by \evo (as mentioned in Section~\ref{sec:test_execution_cost}).

Finally, it is important to highlight that in our experiments we do not track the energy consumption measurement  of the Java Virtual Machine (JVM). This is because different JVM versions have different energy consumption profiles, depending also on the various configuration parameters, such as those concerning garbage collection. Furthermore, our objective in this study is to explore the energy consumption related to the test generation processes without getting into details regarding the JVM configuration, which is outside the scope of this paper. We refer the interested reader to a recent study on the energy consumption of JVMs~\cite{DBLP:conf/esem/OurnaniBRRP21}.

%% file: sections/validity.tex
Threats to \emph{construction validity} of an experiment assess whether the experimental design is concordant with the theory. We used a tool based on Intel's RAPL to quantify energy consumption, branch coverage for measuring code coverage, and cyclomatic complexity to assess code complexity. All metrics are widely adopted in the literature.

Threats to \emph{internal validity} consist of the elements in the experimental design that could impact the observed results. A critical threat relates to the randomness underlying test generation strategies implemented in \evo and energy measurements performed with \jjx. To minimize random fluctuations, we repeated test case generation and the related energy measurements 30 times and used robust statistical methods to compare energy consumption. In the case of per-project test generation, we performed 10 replicas to lower the required computational burden. In this case, to avoid possible threats, we performed qualitative and not quantitative analyses. Another potential threat could originate from the choice of considering only a single test generation tool because it could bias our results. Choosing a different test generation tool could result in different absolute values in, for instance, Figure~\ref{fig:energy_boxplot}. However, by exclusively using \evo, we can effectively minimize potential confounding factors and ensure a fair comparison of test generation algorithms in line with our research questions.

Threats to \emph{conclusion validity} may occur when the conclusions are not consistently derived from the experimental results. 
To support our conclusions,  we utilized rigorous statistical analysis along with repeated executions. We used non-parametric tests to compare distributions, as parametric tests rely on strict assumptions such as variance homogeneity. Furthermore, to account for multiple hypothesis testing, we adopted  the conservative Benjamini-Hochberg procedure  with a significance level of  5\%.

Threats to \emph{external validity} account for the generalization of the results. In our analysis, we only considered a single test generation tool for Java. However, Java is among the most popular programming languages\footnote{https://spectrum.ieee.org/top-programming-languages-2024} and its analysis is significant to the testing community. Moreover, \evo implements different test generation strategies making it a good candidate to check for differences among the various algorithms.
Generalization of the results reported should be taken with care, and future analysis of programming languages and test generation tools could further support our conclusions.
The search budget is another potential source of external validity. We used a budget of 180 seconds, larger than the 120 seconds  used in other large-scale test case
generation experiments~\cite{DBLP:journals/infsof/PanichellaKT18}. 

%% file: sections/conclusions.tex
In this article, we conducted an extended empirical evaluation of the energy consumption associated with automatic test case generation. 
In particular, we compared the energy consumption of five different algorithms implemented in the tool \evo, 
a widely used test generation tool for Java. 
Moreover, we assessed energy usage of test case execution by analyzing two key aspects: the variances among the tests generated by the considered \evo algorithms, and
the differences between automatically generated and manually written tests.

Our study showed varying energy consumption levels among the different test case generation algorithms considered. Moreover, we identified the cyclomatic complexity of the class under test as a key driving factor (Figure~\ref{fig:energy_boxplot}). In particular, Java classes with high cyclomatic complexity tend to require more energy, mainly when time budget is consumed and full coverage is not achieved. Among the algorithms we considered, we found that \nsga and \mupluslambda are the most energy demanding, especially when the cyclomatic complexity is high and complete code coverage is not achieved. 

The analysis of the energy consumption at the method level indicates that there is a significant variation in the energy contribution of individual methods. We have identified a small subset of high energy consuming methods, which are mainly associated with low-level operations such as thread management and I/O operation. 
Methods like 
\texttt{java.lang.Thread.dumpThreads} and \texttt{ jdk.internal.misc.Unsafe.unpark} are consistently among the top consuming methods (Table~\ref{tab:top_consuming_methods_freq}). 
These methods are related to the multi-threaded nature of \evo, which relies on threads to manage the execution of test cases. 
In general, energy analysis at the method level can be combined with recent studies that assess the energy usage of Java libraries, e.g.~\cite{ournani2021comparing} and~\cite{hasan2016energy}, to gain insight into how to reduce the energy impact of test case generation. For instance, the method \texttt{java.util.HashMap.putVal} is among the top consuming methods~(Table~\ref{tab:top_consuming_methods_freq}). Analysis of the Java collection classes highlights that the energy consumed by HashMap insertion varies depending on the specific implementation~\cite{hasan2016energy}. Using a different HashMap implementation in \evo could potentially impact the energy footprint of the automatic test case generation process. 
Interestingly, we also identified a few \evo methods, highlighted in bold in Table~\ref{tab:top_consuming_methods_freq}, which indicate potential optimization can be applied to the internal mechanisms of \evo.

Focusing on the energy consumed by the execution of the tests, it becomes evident that generating tests requires significantly more energy than running them (Figure~\ref{fig:generate_tests_per_project_summary_boxplot}). Moreover, comparing the execution of automatically generated and manually written test suites did not highlight major differences in energy consumption (Figure~\ref{fig:manual_vs_auto_dotplot}). 
However, we observed that some manually written tests are quite complex and energy demanding while achieving lower levels of coverage when compared to more energy efficient automatically generated tests. A deeper evaluation should be performed, but these results suggest that simpler tests require less energy while maintaining comparable coverage levels.

This is an exploratory study and could be extended in various directions in future work. 
The study is based on \evo and mainly focuses on search based software testing of Java programs. 
However, other approaches should be considered from the energy point of view. 
Moreover, other programming languages should be considered in future studies. The work from which we collected our dataset~\cite{DBLP:conf/icse/GruberRPSM024} includes a large set of Python projects with both manually written and automatically generated tests. 
Extending our analysis to this dataset would be very interesting if suitable tools that allow for fine grained energy analysis (like \jjx) become available. 
However, tools like  CodeCarbon~\cite{benoit_courty_2024_11171501} and pyJoules~\cite{fieni2024powerapi} can be adapted to our study in a future extension to evaluate the energy consumption of Python code.
Method level analysis is a promising approach to identify energy leaks in software. The analysis presented is still preliminary and should be extended by analyzing call sequences in a future extension. Looking further ahead, we could consider an energy aware version of \evo, which would be particularly interesting but challenging. This is because energy improvements need to be related to performance, a key feature in test generation, and it would not be trivial to reduce energy consumption while maintaining comparable performance levels. 
The comparison of manually written and automatically generated tests suggests potential  development opportunities. The concept of energy debt~\cite{couto2020energy} has been recently introduced as a metric to capture the energy consumption of software execution, leading to tools like E-Debitum~\cite{maia2020debitum} able to analyze software and identify candidate energy flaws.  E-Debitum targets Java applications for Android, but the same approach could be applied in software testing to envision a tool that supports testers in writing energy aware tests.